\documentclass[prr,twocolumn,aps,amsmath,amssymb,superscriptaddress,floatfix,longbibliography,nofootinbib]{revtex4-1}

\usepackage{latexsym}
\usepackage{amssymb}
\usepackage{graphicx}
\usepackage{amsmath}
\usepackage{bm}
\usepackage{verbatim}
\usepackage{mathrsfs}
\usepackage{extarrows}
\usepackage{comment}
\usepackage{mathtools,slashed}
\usepackage{soul}
\usepackage[toc,page]{appendix}
\usepackage[vcentermath]{youngtab}
\usepackage{multirow}
\usepackage{atbegshi,picture}
\usepackage{lipsum}
\usepackage{amsthm}
\usepackage[colorlinks=true,linktoc=page,linkcolor=blue,citecolor=blue]{hyperref}

\def\:={\,\raisebox{0.85pt}{.}\hspace{-2.78pt}\raisebox{2.85pt}{.}\!\!=\,}
\def\=:{\,=\!\!\raisebox{0.85pt}{.}\hspace{-2.78pt}\raisebox{2.85pt}{.}\,}

\begin{document}

\title{Lower bounds for adiabatic quantum algorithms by quantum speed limits}

\author{Jyong-Hao Chen}
\email{jhchen@lorentz.leidenuniv.nl}
\affiliation{Instituut-Lorentz, Universiteit Leiden, P.O.\ Box 9506, 2300 RA Leiden, The Netherlands}

\date{\today}    

\begin{abstract}
We introduce a simple framework for estimating lower bounds on the runtime of a broad class of adiabatic quantum algorithms.
The central formula consists of calculating the variance of the final Hamiltonian with respect to the initial state.
After examining adiabatic versions of certain keystone circuit-based quantum algorithms, 
this technique is applied to adiabatic quantum algorithms with undetermined speedup.
In particular, we analytically obtain lower bounds on adiabatic algorithms for finding $k$-clique in random graphs. 
Additionally, for a particular class of Hamiltonian, it is straightforward to prove the equivalence between 
our framework and the conventional approach based on spectral gap analysis.
\end{abstract}

\maketitle

\section{Introduction}
Adiabatic approximation in quantum mechanics has found applications in many fields since it was initially developed in the 1920s
\cite{Born27,Born28}.
Nonetheless, the idea that adiabatic approximation may serve as a basis for quantum computing
did not arise until the last two decades \cite{Farhi00,Farhi01_science}.
It has been proved \cite{Aharonov07,Mizel07} that adiabatic quantum computation is computationally equivalent to the standard circuit-based quantum computation \cite{Feynman85,Nielsen10}.	

An adiabatic quantum algorithm (AQA) consists of three components \cite{Choi10,Albash18}:
(i)
a {\it problem} (or {\it final}) {\it Hamiltonian} $H^{\,}_{1}$ whose ground state encodes solutions to the problem;
(ii)
a {\it driver} (or {\it initial}) {\it Hamiltonian} $H^{\,}_{0}$ 
who does not commute with the final Hamiltonian $H^{\,}_{1}$
and whose ground state is known;
and (iii) a timing {\it schedule} $\lambda(t)\in[0,1]$, a time-dependent 
strictly increasing function
that interpolates between the initial Hamiltonian and the final Hamiltonian.
The full time-dependent Hamiltonian $H^{\,}_{\lambda(t)}$ of an AQA is then defined as
\begin{align}
H^{\,}_{\lambda(t)}
&\:=(1-\lambda(t))H^{\,}_{0}+\lambda(t)H^{\,}_{1}
\label{eq: define generic adiabatic Hamiltonian}
\end{align}
with boundary conditions $\lambda(0)=0$ and $\lambda(T)=1$, where $T$ is the runtime of the algorithm.

An AQA is executed by preparing a quantum system at the ground state of the initial Hamiltonian $H^{\,}_{0}$ and then letting it evolve. 
The adiabatic theorem \cite{Born27,Born28,Kato50,Messiah14} guarantees that 
the final state of time evolution 
remains close to 
the instantaneous ground state of the problem Hamiltonian $H^{\,}_{1}$,
provided that the system evolves slowly enough,
or, equivalently, the runtime must be long enough.
A typical sufficient condition for the runtime $T$ is that it must scale 
polynomially with an inverse
of the minimum spectral gap
(the difference between the two smallest energy eigenvalues)
of the full Hamiltonian $H^{\,}_{\lambda}$ \cite{Messiah14,Jansen07,Albash18}.
Nevertheless, analytical calculation of spectral gap is possible only for relatively simple Hamiltonians, 
whereas performing numerical calculations is limited to small system size.

The main contribution of this work is to present 
a more elementary, alternative method to the conventional spectral gap analysis
for estimating the runtime of adiabatic quantum algorithms.
In contrast to the well-known gap conditions \cite{Messiah14,Jansen07,Albash18}, 
which provide {\it sufficient} conditions for runtime,
our formula provides a 
lower bound (or {\it necessary} condition).
More significantly, our technique can be exploited to analytically obtain a lower bound on runtime 
for an adiabatic quantum algorithm with undetermined speedup,
i.e., 
the adiabatic algorithm of Childs {\it et al.}~\cite{Childs02} for finding $k$-clique in random graphs, 
where the conventional method based on spectral gap is unable to do so.

This paper is organized as follows.
In Sec.\ \ref{sec: formalism}, 
we derive the main formula for estimating a {\it necessary} runtime using the quantum uncertainty of the final Hamiltonian with respect to the initial state.
We shall clarify the condition for which the obtained necessary runtime can be interpreted as a {\it lower bound}
(in the sense of computational complexity) on the runtime of AQAs.
After examining our formula with adiabatic versions \cite{Das02,Wei06,Hen14,Farhi01_science,Roland02}
of certain keystone quantum algorithms \cite{Deutsch92,Bernstein93,Bernstein97,Grover96,Grover97} 
in Sec.\ \ref{sec: Final Hamiltonian of type-I} and Sec.\ \ref{sec: Final Hamiltonian of type-II},
our technique is applied in Sec.\ \ref{sec: Adiabatic algorithm for finding k-clique} to analytically obtain lower bounds for adiabatic algorithms of 
finding $k$-clique in random graphs.
Section \ref{sec: Discussion} 
compares our approach to conventional spectral gap analysis and relevant prior works.
Finally, conclusions are presented in Sec.\ \ref{sec: Concluding remarks}.

\section{Formalism}
\label{sec: formalism}
Consider a quantum system characterized by a time-dependent Hamiltonian $H^{\,}_{\lambda}$
with $\lambda=\lambda(t)$ 
being a function of time $t$.
For each $\lambda,$ 
the instantaneous ground state $|\Phi^{\,}_{\lambda}\rangle$ is 
the solution to the eigenvalue problem
$
H^{\,}_{\lambda}|\Phi^{\,}_{\lambda}\rangle=E^{\,}_{\mathrm{GS},\lambda}|\Phi^{\,}_{\lambda}\rangle,
$
where $E^{\,}_{\mathrm{GS},\lambda}$ is the ground state energy. 
On the other hand, the actual dynamics of the quantum system is described by the physical (or time-evolved) state $|\Psi^{\,}_{\lambda}\rangle$, 
which is the solution to the scaled time-dependent Schr\"odinger equation
$
\mathrm{i}\Gamma\partial^{\,}_{\lambda}|\Psi^{\,}_{\lambda}\rangle
=
H^{\,}_{\lambda}|\Psi^{\,}_{\lambda}\rangle
$
with initial condition $|\Psi^{\,}_{0}\rangle=|\Phi^{\,}_{0}\rangle,$
where $\Gamma\:=\partial^{\,}_{t}\lambda(t)$ is the driving rate.
Mathematically, adiabaticity can be quantified by the fidelity between the physical state
$|\Psi^{\,}_{\lambda}\rangle$ and the instantaneous ground state $|\Phi^{\,}_{\lambda}\rangle$, 
namely, the {\it adiabatic fidelity}
$
\mathcal{F}(\lambda)\:=|\langle\Phi^{\,}_{\lambda}|\Psi^{\,}_{\lambda}\rangle|^{2}.
$
In essence, the quantum adiabatic theorem \cite{Born27,Born28,Kato50,Messiah14} 
ensures the following condition
\begin{align}
1-\mathcal{F}(\lambda)\leq\epsilon,
\label{eq: adiabatic condition with allowance}
\end{align}
provided that the driving rate $\Gamma$ is small enough for any given allowance $\epsilon\in[0,1].$

Building upon the seminal work of Ref.\ \cite{Lychkovskiy17},
one of the main results of Ref.\ \cite{Chen21} is the following inequality
that ties the adiabatic fidelity
$\mathcal{F}(\lambda)$, 
the overlap of instantaneous ground states $\mathcal{C}(\lambda)\:=|\langle\Phi^{\,}_{\lambda}|\Phi^{\,}_{0}\rangle|^{2}$, 
and the Bures angle for physical states $\theta(\lambda)\:=\arccos\left(|\langle\Psi^{\,}_{\lambda}|\Psi^{\,}_{0}\rangle|\right)$
through
\begin{align}
|\mathcal{F}(\lambda)-\mathcal{C}(\lambda)|
\quad\leq\quad
\sin\theta(\lambda)
\quad\leq\quad
\sin
\widetilde{\mathcal{R}}(\lambda),
\label{eq: the main inequality of the previous work}
\end{align}
where 
$\widetilde{\mathcal{R}}(\lambda)\:=
\min\left(
\mathcal{R}(\lambda),\frac{\pi}{2}
\right)$,
\begin{subequations}
\label{eq: define the function R}
\begin{align}
&\mathcal{R}(\lambda)\:=\int^{\lambda}_{0}
\frac{\mathrm{d}\lambda'}{|\Gamma(\lambda')|}
\delta E^{\,}_{0}(\lambda'),
\\
&
\delta E^{\,}_{0}(\lambda)\:=
\sqrt{
\langle H^{2}_{\lambda}\rangle^{\,}_{0}
-
\langle H^{\,}_{\lambda} \rangle^{2}_{0}
}.
\end{align}
\end{subequations}
Here, $\langle \cdots\rangle^{\,}_{0}\:=\langle\Psi^{\,}_{0}|\cdots|\Psi^{\,}_{0}\rangle.$
The second inequality in Eq.\ (\ref{eq: the main inequality of the previous work}),
i.e., 
$\theta(\lambda)
\leq
\widetilde{\mathcal{R}}(\lambda),$ sets an upper bound on the Bures angle of physical states and is dubbed as
{\it quantum speed limit} \cite{Mandelstam45,Aharonov90,Anandan90,Vaidman92,Uhlmann92,Pfeifer93a,Deffner17}.
This name is suggested by the fact that 
since the Bures angle $\theta(\lambda)\in[0,\pi/2]$ is a measure of distance, 
the quantum uncertainty 
$\delta E^{\,}_{0}(\lambda)$ signifies speed.

In this work, we explore a union of
the adiabatic condition [Eq.\ (\ref{eq: adiabatic condition with allowance})] 
and the inequality of adiabatic fidelity [Eq.\ (\ref{eq: the main inequality of the previous work})]
at the end of the time evolution of AQAs, $t=T$ [i.e., when $\lambda(t=T)=1$].
Combining Eq.\ (\ref{eq: adiabatic condition with allowance}) with Eq.\ (\ref{eq: the main inequality of the previous work}) 
leads to the following inequality that must be satisfied by the runtime $T$:
\begin{align}
T\geq\frac{\arcsin\left(\max\left(1-\epsilon-\mathcal{C}(1),0\right)\right)}{\mathcal{R}(1)/T}.
\label{eq: main inequality original form}
\end{align}
This inequality may be interpreted as a {\it necessary condition} obeyed by the runtime $T$ of adiabatic quantum evolution.
\begin{proof}
We begin with rewriting the inequality [Eq.\ (\ref{eq: the main inequality of the previous work})]
into the following form:
\begin{align}
\mathcal{C}(\lambda)-\sin\widetilde{\mathcal{R}}(\lambda)
\leq
\mathcal{F}(\lambda)
\leq
\mathcal{C}(\lambda)+\sin\widetilde{\mathcal{R}}(\lambda).
\end{align}
This together with Eq.\ (\ref{eq: adiabatic condition with allowance}) yields
$
1-\epsilon
\;\leq\;
\mathcal{F}(\lambda)
\;\leq\;
\mathcal{C}(\lambda)+\sin\widetilde{\mathcal{R}}(\lambda),
$
which can be expressed as
\begin{align}
\max\left(1-\epsilon-\mathcal{C}(\lambda),0\right)
\quad\leq\quad
\sin\widetilde{\mathcal{R}}(\lambda),
\end{align}
since $\sin\widetilde{\mathcal{R}}(\lambda)\geq0$ by definition.
Now, because $\sin x$ is a monotonically increasing function for $x\in[0,\pi/2]$, the inequality above can be inverted to obtain
\begin{align}
\arcsin\left(\max\left(1-\epsilon-\mathcal{C}(\lambda),0\right)\right)
\leq
\widetilde{\mathcal{R}}(\lambda)
\leq
\mathcal{R}(\lambda),
\end{align}
where we have used $\widetilde{\mathcal{R}}(\lambda)\leq\mathcal{R}(\lambda)$ to obtain the last inequality.
Upon taking $\lambda(t=T)=1,$ the proof of Eq.\ (\ref{eq: main inequality original form}) is therefore complete.
\end{proof}

Notice that inequality Eq.\ (\ref{eq: main inequality original form}) is applicable to Hamiltonians of any given form. 
Next, we shall specifically utilize inequality Eq.\ (\ref{eq: main inequality original form}) for 
AQAs having a full Hamiltonian $H^{\,}_{\lambda(t)}$ defined in Eq.\ (\ref{eq: define generic adiabatic Hamiltonian}).
If so,
the function $\mathcal{R}(\lambda(t))$ per Eqs.\ (\ref{eq: define the function R}) and (\ref{eq: define generic adiabatic Hamiltonian})
reads
\begin{subequations}
\label{eq: R function when restricted to AQA}
\begin{align}
&\mathcal{R}(\lambda(t))
=
t
\overline{\lambda(t)}\,
\delta V^{\,}_{n},
\\
&
\delta V^{\,}_{n}\:=
\sqrt{
\langle H^{2}_{1}\rangle^{\,}_{0}
-
\langle H^{\,}_{1}\rangle^{2}_{0}
}.
\label{eq: uncertainty of driving term final form}
\end{align}
\end{subequations}
Here, $\delta V^{\,}_{n}$
is the quantum uncertainty of the final Hamiltonian $H^{\,}_{1}$ with respect to the initial ground state $|\Phi^{\,}_{0}\rangle$,
and $\overline{\lambda(t)}\:=t^{-1}\int^{t}_{0}\mathrm{d}t'
\lambda(t')$ 
is the time average of the schedule function $\lambda(t).$
The subscript $n$ emphasizes that the quantum uncertainty $\delta V^{\,}_{n}$ often varies with the problem size (or number of qubits) $n$.
Hence, with the help of Eq.\ (\ref{eq: R function when restricted to AQA}),
inequality Eq.\ (\ref{eq: main inequality original form}) can be written as
\begin{align}
T&\geq\frac{\arcsin\left(\max\left(1-\epsilon-\mathcal{C}(1),0\right)\right)}{\overline{\lambda(T)}\,\delta V^{\,}_{n}}.
\label{eq: lower bound on adiabatic runtime}
\end{align}
Now, since the numerator $\arcsin\left(\max\left(1-\epsilon-\mathcal{C}(1),0\right)\right)$ is $\mathcal{O}(1)$,\footnote{Throughout this paper, 
we use $\mathcal{O}(\cdot)$ to indicate ``order of.''}
and because $0<\overline{\lambda(T)}\leq 1$ (as a consequence of $\lambda(t)\in[0,1]$ by definition),
we shall just concentrate on the asymptotic behavior of $\delta V^{\,}_{n}$ [Eq.\ (\ref{eq: uncertainty of driving term final form})]
for the purpose of performing asymptotic analysis using 
Eq.\ (\ref{eq: lower bound on adiabatic runtime}):
\begin{align}
T\geq T^{\,}_{\mathrm{inf}},\quad T^{\,}_{\mathrm{inf}}\:=\mathcal{O}\left(1/\delta V^{\,}_{n}\right).
\label{eq: lower bound on adiabatic runtime asymptotic}
\end{align}
Observe that 
estimating a necessary runtime $T^{\,}_{\mathrm{inf}}$
using Eq.\ (\ref{eq: lower bound on adiabatic runtime asymptotic})
requires only two ingredients, i.e.,
the initial state $|\Phi^{\,}_{0}\rangle$ and 
the final Hamiltonian $H^{\,}_{1}$.
Although the schedule function $\lambda(t)$ is one of the three ingredients for specifying an AQA, 
our formula [Eq.\ (\ref{eq: lower bound on adiabatic runtime asymptotic})] for determining a necessary runtime does not depend on 
any particular form of $\lambda(t)$.
This is unlike the conventional approach of spectral gap analysis \cite{Jansen07} where an explicit form of $\lambda(t)$ must be specified;
given an initial Hamiltonian and a final Hamiltonian, 
different forms of $\lambda(t)$ could result in different estimations of asymptotic forms of runtime.
However, it is not possible for those runtime estimates using spectral gap analysis to be smaller than the necessary runtime determined by 
Eq.\ (\ref{eq: lower bound on adiabatic runtime asymptotic}). 
In other words,
for an AQA specified by an initial state and a final Hamiltonian, one can use 
Eq.\ (\ref{eq: lower bound on adiabatic runtime asymptotic}) to obtain a necessary runtime.
This necessary runtime might be saturated by choosing a particular schedule function or a different initial Hamiltonian.

Although the inequality Eq.\ (\ref{eq: lower bound on adiabatic runtime asymptotic}) is established as a universal necessary condition for the runtime of any AQA, 
interpreting the necessary runtime $T^{\,}_{\mathrm{inf}}$ [Eq.\ (\ref{eq: lower bound on adiabatic runtime asymptotic})] as a {\it lower bound} 
on time complexity only applies to those AQAs whose quantum uncertainty $\delta V^{\,}_{n}$ does not diverge when $n$ goes to infinity,
namely, if
\begin{align}
\lim^{\,}_{n\to\infty}
1/\delta V^{\,}_{n}\,\neq0.
\label{eq: final Hamiltonian of projector like condition}
\end{align}
Otherwise, one would incorrectly deduce that the required runtime decreases as the input size $n$ increases. 
The asymptotic property [Eq.\ (\ref{eq: final Hamiltonian of projector like condition})] can be
fulfilled by a wide class of final Hamiltonians $H^{\,}_{1}$ that meets the following condition\footnote{
Throughout this paper, the notation ``$\overset{!}{=}$'' 
is used to indicate the case where both sides of an equation 
are posited to be equal. 
}
(up to a redefinition\footnote{For instance, if $H^{2}_{1}=-H^{\,}_{1},$ one can define a new Hamiltonian $H^{\prime}_{1}=\mathbb{I}+H^{\,}_{1}$ 
so that ${H}^{\prime2}_{1}={H}^{\prime}_{1}$ holds.
} of $H^{\,}_{1}$):
\begin{align}
\langle H^{2}_{1}\rangle^{\,}_{0}
\overset{!}{=}
\langle H^{\,}_{1}\rangle^{\,}_{0}.
\label{eq: condition for using our formalism}
\end{align}
To see this, imposing the {\it moments condition} [Eq.\ (\ref{eq: condition for using our formalism})] on the definition of quantum uncertainty $\delta V^{\,}_{n}$ 
[Eq.\ (\ref{eq: uncertainty of driving term final form})] yields
$
\delta V^{\,}_{n}=
\sqrt{\langle H^{\,}_{1}\rangle^{\,}_{0}\left(1-\langle H^{\,}_{1}\rangle^{\,}_{0}\right)}.
$
This quantum uncertainty satisfies the desired asymptotic property [Eq.\ (\ref{eq: final Hamiltonian of projector like condition})] since the expectation value 
$\langle H^{\,}_{1}\rangle^{\,}_{0}$ 
is bounded,  
$0\leq\langle H^{\,}_{1}\rangle^{\,}_{0}\leq 1,$
which is a consequence of the moments condition [Eq.\ (\ref{eq: condition for using our formalism})] together with the Cauchy-Schwarz inequality.
If the moments condition [Eq.\ (\ref{eq: condition for using our formalism})] holds, 
the corresponding 
lower bound on runtime [Eq.\ (\ref{eq: lower bound on adiabatic runtime asymptotic})] reads
\begin{align}
T^{\,}_{\mathrm{inf}}=
\mathcal{O}
\left(
1/\delta V^{\,}_{n}
\right),
\quad
\delta V^{\,}_{n}=
\sqrt{\langle H^{\,}_{1}\rangle^{\,}_{0}\left(1-\langle H^{\,}_{1}\rangle^{\,}_{0}\right)}.
\label{eq: minimal runtime for projector Hamiltonian}
\end{align}

A stronger condition than the moments condition [Eq.\ (\ref{eq: condition for using our formalism})] is when the final Hamiltonian is a projector,
namely, if $H^{2}_{1}\overset{!}{=}H^{\,}_{1}$. 
As we shall see shortly, 
the adiabatic version of Deutsch-Jozsa algorithm \cite{Das02,Wei06}, Bernstein-Vazirani algorithm \cite{Hen14}, and Grover search \cite{Farhi01_science,Roland02}
all can be implemented using a projector final Hamiltonian.
A representative final Hamiltonian that does not satisfy the asymptotic property [Eq.\ (\ref{eq: final Hamiltonian of projector like condition})]
is the one containing Ising terms, 
whose quantum uncertainty scales as $\delta V^{\,}_{n}\propto \sqrt{n}$ (see Appendix \ref{sec: App 1}).

In what follows, we shall focus on AQAs that satisfy the moments condition [Eq.\ (\ref{eq: condition for using our formalism})],
and apply formula Eq.\ (\ref{eq: minimal runtime for projector Hamiltonian}) to obtain a lower bound on runtime.
For this purpose, we need to specify what kind of initial state $|\Phi^{\,}_{0}\rangle$ and final Hamiltonian $H^{\,}_{1}$ we wish to use.
For an AQA consisting of $n$ qubits, the initial state $|\Phi^{\,}_{0}\rangle$ can often be chosen as the uniform superposition of 
all basis states of the $2^{n}$-dimensional Hilbert space in the computational basis,
\begin{align}
|\Phi^{\,}_{0}\rangle=\frac{1}{\sqrt{2^n}}\sum^{\,}_{z\in\{0,1\}^{n}}|z\rangle
\=:|+\rangle^{\otimes n}.
\label{eq: initial state with uniform superposition}
\end{align}
Here, 
$|z\rangle=|z^{\,}_{0}\rangle\otimes|z^{\,}_{1}\rangle\otimes\cdots\otimes|z^{\,}_{n-1}\rangle$ with each $z^{\,}_{i}\in\{0,1\}$,
$\{0,1\}^n$ is the set of $2^{n}$ possible $n$-bit binary strings,
and $|\pm\rangle=(|0\rangle\pm|1\rangle)/\sqrt{2}.$
As for the final Hamiltonian $H^{\,}_{1}$,
the following two types of final Hamiltonian satisfying Eq.\ (\ref{eq: condition for using our formalism}) will be discussed in detail:
\begin{itemize}
\item[(i)] Final Hamiltonian with orthogonal projection.

\item[ (ii)] A special class of optimization problems.

\end{itemize}

\section{Final Hamiltonian of type-I}
\label{sec: Final Hamiltonian of type-I}
Let $\Pi^{\,}_{\Phi^{\,}_{1}}$ be an projector that projects onto the eigenspace of the final state $|\Phi^{\,}_{1}\rangle.$
In certain AQAs, their final Hamiltonian $H^{\,}_{1}$ can be implemented by a
complementary projector of $\Pi^{\,}_{\Phi^{\,}_{1}}$
\begin{align}
H^{\,}_{1}=\mathbb{I}-\Pi^{\,}_{\Phi^{\,}_{1}},
\label{Final Hamiltonian of Type I}
\end{align}
where $\mathbb{I}$ is an identity operator.
Adiabatic Deutsch-Jozsa algorithm \cite{Das02,Wei06} and adiabatic Bernstein-Vazirani algorithm \cite{Hen14} are two examples of 
AQAs whose final Hamiltonian has the form shown in Eq.\ (\ref{Final Hamiltonian of Type I}).

\subsection{Example: Adiabatic Deutsch-Jozsa algorithm}
Recall that in the problem of Deutsch-Jozsa \cite{Deutsch92}, we are given a Boolean function $f\colon \{0,1\}^n\to\{0,1\}.$ 
The task is to determine whether $f$ is constant or balanced (i.e., equal number of output of 0's and 1's).
In the AQA proposed by Ref.\ \cite{Das02}, the final state is chosen as (here, $N\equiv 2^n$)
\begin{align}
|\Phi^{\,}_{1}\rangle=
\mu^{\,}_{\mathrm{f}}|0\rangle
+
\frac{1-\mu^{\,}_{\mathrm{f}}}{\sqrt{N}}\sum^{N-1}_{i=1}|i\rangle,
\label{eq: final state adiabatic DJ Das}
\end{align}
where 
$
\mu^{\,}_{\mathrm{f}}\:=N^{-1}
\left|
\sum^{\,}_{z\in\{0,1\}^n}
(-1)^{f(z)}
\right|.
$
If $f$ is constant (respectively, balanced), then $\mu^{\,}_{\mathrm{f}}=1$ (respectively, $\mu^{\,}_{\mathrm{f}}=0$).
The corresponding final Hamiltonian can be constructed as a projector
shown in Eq.\ (\ref{Final Hamiltonian of Type I}) with $\Pi^{\,}_{\Phi^{\,}_{1}}=|\Phi^{\,}_{1}\rangle\langle \Phi^{\,}_{1}|.$
The initial state is chosen as an equal weight superposition of all basis states [Eq.\ (\ref{eq: initial state with uniform superposition})].
In Ref.\ \cite{Das02}, using conventional spectral gap analysis yields the scaling of runtime $T\sim\mathcal{O}(N)$ for $\lambda(t)=\frac{t}{T}$
and $T\sim\mathcal{O}(\sqrt{N})$ for a local adiabatic evolution. 
These results do not reach the well-known \cite{Deutsch92} optimal runtime, i.e., $\mathcal{O}(1)$.
Nevertheless, we shall demonstrate that using the formula Eq.\ (\ref{eq: minimal runtime for projector Hamiltonian}) enables us to 
obtain the lowest runtime found in Ref.\ \cite{Das02}.
We compute the overlap 
$\left|
\langle\Phi^{\,}_{1}|\Phi^{\,}_{0}\rangle
\right|^2$
using Eqs.\ (\ref{eq: initial state with uniform superposition}) and (\ref{eq: final state adiabatic DJ Das}) as
\begin{align}
\left|
\langle\Phi^{\,}_{1}|\Phi^{\,}_{0}\rangle
\right|^2
&=
\frac{1}{N}
\left(
\mu^{\,}_{\mathrm{f}}
+
(1-\mu^{\,}_{\mathrm{f}})
\sqrt{N-1}
\right)^2,
\label{eq: C function DJ Das asymptotic}
\end{align}
and apply it to Eq.\ (\ref{eq: minimal runtime for projector Hamiltonian}) to obtain
$
T^{\,}_{\mathrm{inf}}=\mathcal{O}(\sqrt{N}),
$
which is consistent with the result obtained by Ref.\ \cite{Das02} as mentioned above.

From the general expression of $T^{\,}_{\mathrm{inf}}$ [Eq.\ (\ref{eq: minimal runtime for projector Hamiltonian})],
it is clear that in order to reproduce the correct optimal runtime of the Deutsch-Jozsa algorithm, 
the overlap 
$\left|
\langle\Phi^{\,}_{1}|\Phi^{\,}_{0}\rangle
\right|^2$
should be independent of $N$.
In light of this, the final state [Eq.\ (\ref{eq: final state adiabatic DJ Das})] should be modified. 
Exactly this modification has been carried out in Ref.\ \cite{Wei06}
by replacing the final state of Eq.\ (\ref{eq: final state adiabatic DJ Das}) 
with the one which is more symmetric in amplitudes:
\begin{align}
|\Phi^{\,}_{1}\rangle=
\frac{\mu^{\,}_{\mathrm{f}}}{\sqrt{N/2}}
\sum^{N/2-1}_{i=0}|2i\rangle
+
\frac{1-\mu^{\,}_{\mathrm{f}}}{\sqrt{N/2}}
\sum^{N/2-1}_{i=0}|2i+1\rangle.
\label{eq: final state adiabatic DJ WY}
\end{align}
Upon using Eqs.\ (\ref{eq: initial state with uniform superposition}) and (\ref{eq: final state adiabatic DJ WY}), the overlap $\left|
\langle\Phi^{\,}_{1}|\Phi^{\,}_{0}\rangle
\right|^2$ reads
\begin{align}
\left|
\langle\Phi^{\,}_{1}|\Phi^{\,}_{0}\rangle
\right|^2
&=
\frac{1}{N}
\frac{1}{N/2}
\left(
\mu^{\,}_{\mathrm{f}}\frac{N}{2}
+
(1-\mu^{\,}_{\mathrm{f}})\frac{N}{2}
\right)^2
=
\frac{1}{2}.
\label{eq: C function DJ WY asymptotic}
\end{align}
This result should be compared with that of Eq.\ (\ref{eq: C function DJ Das asymptotic}).
Bringing Eq.\ (\ref{eq: C function DJ WY asymptotic}) to Eq.\ (\ref{eq: minimal runtime for projector Hamiltonian}) yields
$
T^{\,}_{\mathrm{inf}}=\mathcal{O}(1).
$
Therefore, the optimal runtime for the Deutsch-Jozsa problem is attained.
In contrast to the usual approach of estimating runtime via spectral gap as done in Refs.\ \cite{Das02,Wei06}, 
our formalism clearly illustrates why the algorithm of Ref.\ \cite{Wei06} is superior to that of Ref.\ \cite{Das02}.

\subsection{Example: Adiabatic Bernstein-Vazirani algorithm}
As the second example for the final Hamiltonian of Type-I [Eq.\ (\ref{Final Hamiltonian of Type I})], 
we consider an adiabatic algorithm solving the Bernstein-Vazirani problem.
Recall that in the Bernstein-Vazirani problem \cite{Bernstein93,Bernstein97}, one is given an oracle that evaluates the function
$f^{\,}_{\lambda}: \{0,1\}^n\mapsto\{0,1\}$ with
$
f^{\,}_{\lambda}(z)
=
\sum^{n-1}_{i=0}
z^{\,}_{i}s^{\,}_{i}
\bmod 2.
$
The task is to find the unknown $n$-bit binary string $s\in\{0,1\}^n$ using as few queries of the function $f^{\,}_{\lambda}$ as possible.
The adiabatic algorithm proposed by Ref.\ \cite{Hen14} is the following. 
First, notice that the oracle function $f^{\,}_{\lambda}$ can be encoded in a projector final Hamiltonian as shown in Eq.\ (\ref{Final Hamiltonian of Type I})
and acting on two subsystems, $\mathrm{A}$ and $\mathrm{B}$, comprising $n$ qubits and 1 qubit, respectively,
\begin{equation}
\begin{split}
&H^{\,}_{1}=
\mathbb{I}^{\,}_{\mathrm{A}}\otimes\mathbb{I}^{\,}_{\mathrm{B}}
-\Pi^{\,}_{\Phi^{\,}_{1}},
\\
&
\Pi^{\,}_{\Phi^{\,}_{1}}=
\sum^{\,}_{z\in\{0,1\}^n}
|z\rangle^{\,}_{\mathrm{A}}\langle z|\otimes
|f^{\,}_{\lambda}(z)\rangle^{\,}_{\mathrm{B}}\langle f^{\,}_{\lambda}(z)|.
\label{eq: final Hamiltonian adiabatic BV}
\end{split}
\end{equation}
The ground state of $H^{\,}_{1}$
is 
$
|\Phi^{\,}_{1}\rangle=
\frac{1}{\sqrt{2^{n}}}
\sum^{\,}_{z\in\{0,1\}^n}
|z\rangle^{\,}_{\mathrm{A}}\otimes|f^{\,}_{\lambda}(z)\rangle^{\,}_{\mathrm{B}}
$
with eigenvalue $0$.
The initial state is again
the uniform superposition state [Eq.\ (\ref{eq: initial state with uniform superposition})]
$
|\Phi^{\,}_{0}\rangle
=
|+\rangle^{\otimes n}_{\mathrm{A}}
\otimes
|+\rangle^{\,}_{\mathrm{B}}.
$
We refer to Refs.\ \cite{Hen14,Albash18} for further details on the mechanism behind this adiabatic algorithm.
For our purpose, we proceed to use the formula Eq.\ (\ref{eq: minimal runtime for projector Hamiltonian})
to obtain a lower bound on the runtime.
We first note that $H^{\,}_{1}$ defined in Eq.\ (\ref{eq: final Hamiltonian adiabatic BV}) is apparently a projector, $H^{2}_{1}=H^{\,}_{1}$.
The remaining task is to compute 
$\langle\Phi^{\,}_{0}|H^{\,}_{1}|\Phi^{\,}_{0}\rangle$: 
\begin{align}
\langle\Phi^{\,}_{0}|H^{\,}_{1}|\Phi^{\,}_{0}\rangle
&=
1-
\frac{1}{2^n}
\sum^{\,}_{z\in\{0,1\}^{n}}
|\langle+|f^{\,}_{\lambda}(z)\rangle|^{2}
\nonumber\\
&\stackrel{(*)}{=}
1-
\frac{1}{2^n}
\sum^{\,}_{z\in\{0,1\}^{n}}
\left(\frac{1}{\sqrt{2}}\right)^2
=
\frac{1}{2},
\label{eq: first moment BV}
\end{align}
where in $(*)$ we have used the following identity
$|f^{\,}_{\lambda}(z)\rangle^{\,}_{\mathrm{B}}=
\frac{1}{\sqrt{2}}
\left(
|+\rangle^{\,}_{\mathrm{B}}
+
(-1)^{f^{\,}_{\lambda}(z)}
|-\rangle^{\,}_{\mathrm{B}}
\right).
$
Finally, it follows from Eq.\ (\ref{eq: minimal runtime for projector Hamiltonian}) that
$
T^{\,}_{\mathrm{inf}}=\mathcal{O}(1)
$
as expected.

\section{Final Hamiltonian of type-II}
\label{sec: Final Hamiltonian of type-II}
For optimization problems with a cost function $h\colon\{0,1\}^n\to\mathbb{R}$, we seek a minimum of $h.$
In the framework of adiabatic quantum computation \cite{Farhi00,Farhi01_science}, 
a final Hamiltonian can be defined to be diagonal in the computational basis, with cost function $h$
being the diagonal element
\begin{align}
H^{\,}_{1}=\sum^{\,}_{z\in\{0,1\}^n}h(z)|z\rangle\langle z|.
\label{Final Hamiltonian of Type II}
\end{align}
If the uniform superposition of all basis states [Eq.\ (\ref{eq: initial state with uniform superposition})] is chosen as the initial state,
the quantum uncertainty [Eq.\ (\ref{eq: uncertainty of driving term final form})] reads 
$
\delta V^{\,}_{n}=(\overline{h^2}-\overline{h}^2)^{1/2},
$
with the arithmetic average
$
\overline{h}
\:=
2^{-n}
\sum^{\,}_{z}h(z)
$
and
$
\overline{h^2}
\:=
2^{-n}
\sum^{\,}_{z}h^2(z).
$
Furthermore, if we impose the moments condition [Eq.\ (\ref{eq: condition for using our formalism})],
which amounts to requiring
\begin{align}
\overline{h^2}
\overset{!}{=}
\overline{h},
\label{eq: condition for restrcited combinotorial optimization problem}
\end{align}
then 
the lower bound $T^{\,}_{\mathrm{inf}}$ [Eq.\ (\ref{eq: minimal runtime for projector Hamiltonian})]
\begin{align}
T^{\,}_{\mathrm{inf}}
=
\mathcal{O}
\left(
1/\delta V^{\,}_{n}
\right)
\quad
\text{with}
\quad
\delta V^{\,}_{n}=
\sqrt{\overline{h}\left(1-\overline{h}\right)}
\label{eq: lower bound on runtime type II}
\end{align}
is completely determined by the arithmetic average of the cost function,
i.e., $\overline{h}.$
Note that 
$0\leq\overline{h}\leq1$ 
as a result of Cauchy-Schwarz
together with the moments condition [Eq.\ (\ref{eq: condition for restrcited combinotorial optimization problem})].
Note also that the moments condition Eq.\ (\ref{eq: condition for restrcited combinotorial optimization problem}) 
can be expressed using matrix analysis terminology \cite{Bhatia96,Horn12} as $\|H^{\,}_{1}\|^{2}_{\mathrm{F}}\overset{!}{=}\mathrm{tr}H^{\,}_{1}$,
where $\|\cdot\|^{\,}_{\mathrm{F}}$ denotes the Frobenius norm (or Hilbert-Schmidt norm). 
If it is further assumed that the cost function $h$ is non-negative, the moments condition [Eq.\ (\ref{eq: condition for restrcited combinotorial optimization problem})]
can alternatively be written as 
$\|H^{\,}_{1}\|^{2}_{\mathrm{F}}\overset{!}{=}\|H^{\,}_{1}\|^{\,}_{\mathrm{tr}}$, where $\|\cdot\|^{\,}_{\mathrm{tr}}$ represents the trace norm (or nuclear norm).

\subsection*{Example: Adiabatic Grover search}
Taking the unstructured search problem of Grover \cite{Grover96,Grover97} as an example,
it can be formulated as a combinatorial optimization problem with the following cost function \cite{Farhi01_science,Dam01}:
$h(z)=0$ for $M$ marked items and $h(z)=1$ otherwise.
For this case, one immediately finds $\overline{h^2}=\overline{h}=1-M/N
$ 
(here, $N\equiv 2^n$).
It then follows from Eq.\ (\ref{eq: lower bound on runtime type II}) that a lower bound 
$
T^{\,}_{\mathrm{inf}}=\mathcal{O}
(
\sqrt{N/M}
),
$
as expected \cite{Bennett97,Farhi98,Roland02}.
We note that the final Hamiltonian of adiabatic Grover search can be equivalently written as $H^{\,}_{1}=\mathbb{I}-\sum^{\,}_{z\in\mathcal{M}}|z\rangle\langle z|$,
where $\mathcal{M}$ is the space of solution (of size $M$). 
Hence, this final Hamiltonian also belongs to the type-I final Hamiltonian [see Eq.\ (\ref{Final Hamiltonian of Type I})]. 
It is with this expression of projector final Hamiltonian that
Ref.\ \cite{Lychkovskiy18} 
used a formula similar to Eq.\ (\ref{eq: lower bound on adiabatic runtime})
to obtain the optimal runtime for adiabatic Grover search.
For completeness, we present a similar calculation using our formula in Appendix \ref{sec: App 2}.

Thus far, three adiabatic versions of keystone quantum algorithms have been examined.
We shall now apply our technique to the adiabatic algorithm proposed by Childs {\it et al.}~\cite{Childs02} for finding $k$-clique in random graphs.

\section{Adiabatic algorithm for finding $k$-clique}
\label{sec: Adiabatic algorithm for finding k-clique}
Consider a random graph $G$ where every pair of vertices is connected or disconnected with a probability of 1/2.
A {\it clique} is a subgraph of $G$ in which every pair of vertices is connected by an edge.
The problem of finding cliques of $k$ vertices (called {\it $k$-clique}) in a random graph of $n$ vertices 
is an NP-complete problem 
if both $n$ and $k$ are treated as inputs \cite{Karp72,Moore12}.

In the algorithm proposed by Ref.\ \cite{Childs02}, each vertex is associated with a qubit. 
Hence, there are $n$ qubits for a graph of $n$ vertices. 
As before, each qubit state is represented by $|z^{\,}_{i}\rangle$ with $z^{\,}_{i}\in\{0,1\}$ for $i\in\{0,\cdots,n-1\}.$
A vertex $i$ is included into a subgraph of $k$ vertices only if $z^{\,}_{i}=1.$

Although the full Hilbert space is $2^n$-dimensional,
we can focus only on the subspace spanned by quantum states with Hamming weight $k$
since we are only interested in those quantum states that represent cliques of $k$ vertices.
For notational convenience, we denote the set of $n$-bit binary strings of Hamming weight $k$ as
$\mathcal{S}^{\,}_{k}\:=\left\{z\in\{0,1\}^{n}\colon|z|=k\right\},$
where $|z|=\sum^{n-1}_{i=0}z^{\,}_{i}$ is the Hamming weight of the $n$-bit binary string $z.$
The size of $\mathcal{S}^{\,}_{k}$ is $\binom{n}{k}\equiv C(n,k).$
In the subspace generated by binary strings from $\mathcal{S}^{\,}_{k}$,
the initial state can be chosen as a 
Dicke state \cite{Dicke54,Toth07,Bartschi19}
\begin{align}
|\Phi^{\,}_{0}\rangle\:=&
\frac{1}{\sqrt{C(n,k)}}
\sum^{\,}_{z\in\mathcal{S}^{\,}_{k}}|z\rangle.
\label{eq: initial state adiabatic clique in RG}
\end{align}

In Ref.\ \cite{Childs02},
the final Hamiltonian is of the form
shown in Eq.\ (\ref{Final Hamiltonian of Type II})
with the cost function
\begin{align}
h^{\,}_{\mathrm{C}}(z)
\:=
\sum^{n-1}_{i,j=0:i>j}
\left(
1-
G^{\,}_{ij}
\right)z^{\,}_{i}z^{\,}_{j},
\label{eq: final Hamiltonian adiabatic clique in RG eg equation}
\end{align}
where $G^{\,}_{ij}$ with $i,j\in\{0,\cdots,n-1\}$ is the matrix element of the adjacency matrix for a graph $G.$ 
As usual, $G^{\,}_{ij}=1$ if the vertices $i$ and $j$ are connected by an edge;
otherwise, $G^{\,}_{ij}=0.$
Observe that 
the cost function $h^{\,}_{\mathrm{C}}(z)$ takes values in $\{0,1,\cdots,L^{\,}_{k}\}$,
where $L^{\,}_{k}\:=C(k,2)$ is the total number of edges for a graph of $k$ vertices.

We proceed to obtain a lower bound on runtime using Eq.\ (\ref{eq: lower bound on adiabatic runtime asymptotic}) for the algorithm defined above.
Since each $G^{\,}_{ij}\in\{0,1\}$ is a random variable, 
an exact computation for the arithmetic average $\overline{h^{\,}_{\mathrm{C}}}$ and $\overline{h^{2}_{\mathrm{C}}}$
is not possible without knowing an explicit instance of $G^{\,}_{ij}.$
We attempt to consider a ``mean-field'' (or ``randomized'') approach by replacing $G^{\,}_{ij}$ with its (classical) expectation value $\mathbb{E}[G^{\,}_{ij}]=1/2,$
and obtain (see Appendix \ref{sec: App 3})
$\mathbb{E}[\overline{h^{\,}_{\mathrm{C}}}]=L^{\,}_{k}/2$ 
and
$\mathbb{E}[\overline{h^{2}_{\mathrm{C}}}]=(L^{\,}_{k}+1)L^{\,}_{k}/4.$
It follows from Eq.\ (\ref{eq: lower bound on adiabatic runtime asymptotic}) that a lower bound on runtime $T^{\,}_{\mathrm{rand},\mathrm{inf}}$ reads
$T^{\,}_{\mathrm{rand},\mathrm{inf}}=\mathcal{O}(\sqrt{1/L^{\,}_{k}}),$
which is independent of $n.$
It is likely that this $n$-independent lower bound is considerably lower than the sufficient runtime obtained using the spectral gap analysis
(though it is not available).
We also notice that the moments condition 
[Eq.\ (\ref{eq: condition for restrcited combinotorial optimization problem})]
is not met for the cost function defined in 
Eq.\ (\ref{eq: final Hamiltonian adiabatic clique in RG eg equation}).
Nevertheless, we note that the $k$ dependence in $T^{\,}_{\mathrm{rand},\mathrm{inf}}$ seems consistent with the numerical data found in Ref.\ \cite{Childs02},
saying that the median runtime for finding cliques of $k=5$ is longer than that for finding cliques of $k=6$.
Specifically, $T^{\,}_{\mathrm{med}}(k=5)/T^{\,}_{\mathrm{med}}(k=6)=30.87/18.56\approx1.66$ found in Ref.\ \cite{Childs02}, 
whereas our 
result indicates
$T^{\,}_{\mathrm{rand},\mathrm{inf}}(k=5)/T^{\,}_{\mathrm{rand},\mathrm{inf}}(k=6)=\sqrt{15/10}\approx1.22.$

To make better use of our formula Eq.\ (\ref{eq: lower bound on runtime type II}),
we propose the following deformed cost function
\begin{align}
h(z)
\:=
\frac{1}{2}
\left(
1
+
h^{\,}_{\mathrm{C}}(z)
-
\left|
1-
h^{\,}_{\mathrm{C}}(z)
\right|
\right).
\label{eq: final Hamiltonian adiabatic clique in RG eg equation new form}
\end{align}
This amounts to introduce
$h(z)=\min(h^{\,}_{\mathrm{C}}(z),1),$
where $h^{\,}_{\mathrm{C}}(z)$ is defined in Eq.\ (\ref{eq: final Hamiltonian adiabatic clique in RG eg equation}).
In other words, those $h^{\,}_{\mathrm{C}}(z)>1$ are mapped to $h(z)=1.$
Therefore, the deformed cost function 
is Boolean-valued $h(z)\in\{0,1\}.$
This deformed cost function is similar to that of adiabatic Grover search discussed previously.
One finds that the cost function defined in Eq.\ (\ref{eq: final Hamiltonian adiabatic clique in RG eg equation new form}) 
along with the initial state defined in Eq.\ (\ref{eq: initial state adiabatic clique in RG}) satisfies the moments condition 
[Eq.\ (\ref{eq: condition for restrcited combinotorial optimization problem})].
Explicitly,
$\overline{h}
=
\overline{h^2}
=
\left(
1-
M/C(n,k)
\right),
$
which renders 
a lower bound from Eq.\ (\ref{eq: lower bound on runtime type II})
\begin{align}
T^{\,}_{\mathrm{inf}}=\mathcal{O}\left(\sqrt{C(n,k)/M}\right),
\label{eq: lower bound for adiabatic k clique in RG final form asymptotic}
\end{align}
where $M$ is the number of cliques of $k$ vertices.
The result [Eq.\ (\ref{eq: lower bound for adiabatic k clique in RG final form asymptotic})] is
consistent with that of a circuit-based algorithm of Ref.\ \cite{Metwalli20}
in which the time complexity is related to the number of  Grover iterations.
For the special case of $n\gg k$, Eq.\ (\ref{eq: lower bound for adiabatic k clique in RG final form asymptotic}) simplifies
$
T^{\,}_{\mathrm{inf}}=
\mathcal{O}\left(n^{k/2}\right)
$
for 
$
n\gg k.
$
A well-studied case is $k=3,$ i.e., the so-called {\it 3-clique problem} (or {\it triangle-finding} problem);
the resulting lower bound is $T^{\,}_{\mathrm{inf}}=
\mathcal{O}(n^{3/2}).$
The exponent, $3/2$, 
agrees with that found in a 
quantum algorithm \cite{Buhrman05} 
for the triangle-finding problem 
using a plain Grover search.

\section{Discussions}
\label{sec: Discussion}

\subsection{Connection to spectral gap analysis}
We shall now attempt to connect our approach with conventional spectral gap analysis.
An equivalence between the two approaches can be directly shown for a particular class of Hamiltonians 
where both the initial and final Hamiltonians are of the projector form
$H^{\,}_{0}=\mathbb{I}-|\Phi^{\,}_{0}\rangle\langle\Phi^{\,}_{0}|$
and $H^{\,}_{1}=\mathbb{I}-|\Phi^{\,}_{1}\rangle\langle\Phi^{\,}_{1}|$.
It was then proved in Ref.\ \cite{Aharonov07s} that the spectral gap of the full Hamiltonian $H^{\,}_{\lambda}$ [Eq.\ (\ref{eq: define generic adiabatic Hamiltonian})],
denoted as $g(H^{\,}_{\lambda}),$ is bounded from below:
$
g(H^{\,}_{\lambda})\geq|\langle\Phi^{\,}_{1}|\Phi^{\,}_{0}\rangle|\equiv g^{\,}_{\mathrm{min}},
$
where $g^{\,}_{\mathrm{min}}=\min^{\,}_{\lambda\in[0,1]}g(H^{\,}_{\lambda})$ is the minimal spectral gap.
If the schedule function $\lambda(t)$ is chosen simply as $\lambda(t)=t/T,$
spectral gap analysis \cite{Roland02,Jansen07} yields the scaling of runtime
$
T^{\,}_{\mathrm{gap}}\sim
\mathcal{O}(1/g^{2}_{\mathrm{min}}).
$
Furthermore, it is possible to improve \cite{Roland02,Jansen07} the error dependence on the minimal gap 
by adopting a nonlinear schedule function; if so,
one obtains
$
T^{\prime}_{\mathrm{gap}}\sim
\mathcal{O}(1/g^{\,}_{\mathrm{min}}).
$
On the other hand,
using our formalism, we directly obtain 
$
\langle H^{\,}_{1}\rangle^{\,}_{0}
=
\langle H^{2}_{1}\rangle^{\,}_{0}
=1-|\langle\Phi^{\,}_{1}|\Phi^{\,}_{0}\rangle|^2=1-g^{2}_{\mathrm{min}}.
$
Consequently,
a lower bound on runtime follows from Eq.\ (\ref{eq: minimal runtime for projector Hamiltonian}):
$
T^{\,}_{\mathrm{inf}}=\mathcal{O}(1/g^{\,}_{\mathrm{min}}).
$

If the minimal gap $g^{\,}_{\mathrm{min}}=|\langle\Phi^{\,}_{1}|\Phi^{\,}_{0}\rangle|$ is independent of $n$, 
then
the two estimates $T^{\,}_{\mathrm{gap}}$ and $T^{\,}_{\mathrm{inf}}$
are equal: $T^{\,}_{\mathrm{gap}}= T^{\,}_{\mathrm{inf}}=\mathcal{O}(1)$.
This is the case for the adiabatic Deutsch-Jozsa algorithm and the adiabatic Bernstein-Vazirani algorithm
discussed previously.
For the case of adiabatic Grover search, the optimal runtime obtained from the spectral gap analysis 
along with a nonlinear schedule function yields \cite{Roland02}
$T^{\prime}_{\mathrm{gap}}=\mathcal{O}(1/g^{\,}_{\mathrm{min}})$,
which again agrees with our lower bound $T^{\,}_{\mathrm{inf}}=\mathcal{O}(1/g^{\,}_{\mathrm{min}})$.

\subsection{Comparison with prior works}
The performance of adiabatic quantum algorithms has been extensively investigated in a variety of settings. 
Among the existing research on this topic, three groups of studies that utilize quantum speed limits are particularly relevant to our current work.
(i) A new class of quantum speed limits is presented in Refs.\ \cite{Kieu19a,Kieu19b} and applied to adiabatic searches. 
A lower bound on runtime is defined there as the minimum time it takes for the physical state to be orthogonal to the initial state. 
This definition is, however, too restrictive.~\footnote{
Indeed, the final state of most adiabatic quantum algorithms is never orthogonal to the initial state, even though the overlap between the two is typically very small.	
}
Additionally, it is unclear whether the results obtained in Refs.\ \cite{Kieu19a,Kieu19b} are applicable beyond adiabatic searches.

(ii)
While the main result of Ref.\ \cite{Lychkovskiy18} is an inequality similar to Eq.\ (\ref{eq: lower bound on adiabatic runtime}),
the present work makes several significant advances.
Our method for estimating lower bounds is more systematic and rigorous, 
and we carefully clarified the applicability of Eq.\ (\ref{eq: lower bound on adiabatic runtime}). 
In particular, we highlighted the importance of the asymptotic property [Eq.\ (\ref{eq: final Hamiltonian of projector like condition})]
and the irrelevance of the schedule function $\lambda(t)$ in asymptotic analysis. 
Furthermore, we explored a wider range of adiabatic quantum algorithms, including the adiabatic Grover algorithm examined in Ref.\ \cite{Lychkovskiy18} 
as well as several additional adiabatic algorithms. 
Notably, we obtained an analytical result on lower bounds for the adiabatic algorithm of Childs {et al.}~\cite{Childs02} for 
finding $k$-clique in random graphs, which is an algorithm with undetermined quantum speedup.

(iii)
It is also worth noting that
quantum speed limits and the formalism of shortcuts to adiabaticity \cite{Demirplak03,Demirplak05,Berry09,Chen10,Torrontegui13,Campo13,Jarzynski13,Odelin19} 
are combined in Ref.\ \cite{Suzuki20} 
to investigate the performance of adiabatic quantum computation.
However, the purpose and scope of Ref.\ \cite{Suzuki20} is significantly different from ours.

\section{Concluding remarks}
\label{sec: Concluding remarks}
We have shown that for a wide class of adiabatic quantum algorithms (AQAs)
in which 
the asymptotic property presented in
Eq.\ (\ref{eq: final Hamiltonian of projector like condition}) holds, 
nontrivial lower bounds on the runtime of adiabatic algorithms can be estimated by calculating 
the quantum uncertainty of the final Hamiltonian with respect to the initial state
[see Eq.\ (\ref{eq: lower bound on adiabatic runtime asymptotic})].
A runtime estimation obtained by conventional spectral gap analysis is by no means smaller than the necessary runtime derived using our method.
The reason is that our formula provides a necessary condition that must be obeyed by the runtime of adiabatic quantum evolution.
Choosing a specific schedule function or initial Hamiltonian may cause the necessary runtime to be saturated.

Our findings may shed new light on the design of new adiabatic quantum algorithms.
For instance, if a potential quantum speedup of an adiabatic quantum algorithm 
is undetermined, 
one may attempt a deformation of the original adiabatic algorithm to fulfill the moments condition 
[Eq.\ (\ref{eq: condition for using our formalism})]
and then apply 
Eq.\ (\ref{eq: minimal runtime for projector Hamiltonian})
to obtain a lower bound on runtime.
The runtime of the deformed algorithm thus obtained may be utilized to estimate the spectral gap (and hence, a sufficient runtime) 
of the original algorithm 
by means of matrix inequalities such as Weyl's inequalities \cite{Bhatia96,Horn12}.
For optimization problems, it is important to emphasize that the lower bound formula [Eq.\ (\ref{eq: lower bound on runtime type II})]
is not valid for every optimization problem but only for those in which the cost function meets the moments condition 
[Eq.\ (\ref{eq: condition for restrcited combinotorial optimization problem})].
Therefore, it would be interesting to investigate further the implications of the moments condition on general Boolean functions 
using techniques from Boolean function analysis
\cite{Donnell14}.
Another future direction would be to derive bounds on the runtime of adiabatic quantum algorithms in open quantum systems.

\section*{Acknowledgments}
This work was funded by the Dutch Research Council (NWO) through
the project Adiabatic Protocols in Extended Quantum Systems (Project No. 680-91-130).




\appendix

\section{Quantum uncertainty of Ising terms}
\label{sec: App 1}

Consider a final Hamiltonian $H^{\,}_{1}$ that consists of Ising terms
$
H^{\,}_{1}=-\sum^{n-1}_{i=0}Z^{\,}_{i}Z^{\,}_{i+1},
$
where $Z^{\,}_{i}$ is the third Pauli matrix acting on the $i$th qubit.
Recall that $Z|+\rangle=|-\rangle$ and $Z|-\rangle=|+\rangle.$
Choosing the initial state $|\Phi^{\,}_{0}\rangle$ as the uniform superposition state [Eq.\ (\ref{eq: initial state with uniform superposition})].
We first compute
\begin{widetext} 
\begin{align}
|\tilde{\Phi}\rangle
&\:=H^{\,}_{1}|\Phi^{\,}_{0}\rangle
=
-\sum^{n-1}_{i=0}Z^{\,}_{i}Z^{\,}_{i+1}|+\rangle^{\otimes n}
=
-\sum^{n-1}_{i=0}
\left(
|-\rangle^{\,}_{i}\otimes|-\rangle^{\,}_{i+1}
\otimes
|+\rangle^{\otimes n-2}
\right).
\end{align} 
We then find
\begin{equation}
\begin{split}
&\langle\Phi^{\,}_{0}|H^{\,}_{1}|\Phi^{\,}_{0}\rangle
=\langle\Phi^{\,}_{0}|\tilde{\Phi}\rangle
=
\langle+|^{\otimes n}\left(
-\sum^{n-1}_{i=0}
\left(
|-\rangle^{\,}_{i}\otimes|-\rangle^{\,}_{i+1}
\otimes
|+\rangle^{\otimes n-2}
\right)
\right)
=0,
\\
&\langle\Phi^{\,}_{0}|H^{2}_{1}|\Phi^{\,}_{0}\rangle
=
\langle\tilde{\Phi}|\tilde{\Phi}\rangle
=
\sum^{n-1}_{i,j=0}
\left(
\langle-|^{\,}_{i}\otimes\langle-|^{\,}_{i+1}
\otimes
\langle+|^{\otimes n-2}
\right)
\left(
|-\rangle^{\,}_{j}\otimes|-\rangle^{\,}_{j+1}
\otimes
|+\rangle^{\otimes n-2}
\right)
=
\sum^{n-1}_{i,j=0}
\delta^{\,}_{ij}
=n.
\end{split}
\end{equation}
\end{widetext} 
It then follows from Eq.\ (\ref{eq: uncertainty of driving term final form}) that the quantum uncertainty reads
$
\delta V^{\,}_{n}=\sqrt{n}.
$

\section{Detail calculation for adiabatic Grover search}
\label{sec: App 2}

Consider a set of $N$ items among which $M<N$ items are marked, 
the goal being to find marked items in minimum time. 
We use $n$ qubits to encode the $N$ items. 
Hence, the Hilbert space is of dimension $N=2^n$. 
In this space, the basis states can be written as
$|i\rangle$ with $i\in\{0,\cdots,N-1\}.$
The desired final state is
$
|\Phi^{\,}_{1}\rangle=\frac{1}{\sqrt{M}}\sum^{\,}_{m\in\mathcal{M}}|m\rangle,
$
and the corresponding final Hamiltonian can be chosen as a projector
$
H^{\,}_{1}=\mathbb{I}-\sum^{\,}_{m\in\mathcal{M}}|m\rangle\langle m|,
$
where $\mathcal{M}$ is the space of solution (of size $M$).
One can easily see that the ground state of $H^{\,}_{1}$ 
is $|\Phi^{\,}_{1}\rangle$ 
with eigenenergy zero. 

The remaining task is to calculate the quantum uncertainty $\delta V^{\,}_{n}$ [Eq.\ (\ref{eq: uncertainty of driving term final form})]
with $|\Phi^{\,}_{0}\rangle$ given by the uniform superposition state [Eq.\ (\ref{eq: initial state with uniform superposition})].
First, we calculate the overlap $|\langle \Phi^{\,}_{1}|\Phi^{\,}_{0}\rangle|^2,$ 
\begin{align}
|\langle \Phi^{\,}_{1}|\Phi^{\,}_{0}\rangle|^{2}
&=
\left|
\left(
\frac{1}{\sqrt{M}}
\sum^{\,}_{m\in\mathcal{M}}\langle m|
\right)
\left(
\frac{1}{\sqrt{N}}
\sum^{N-1}_{i=0}|i\rangle
\right)
\right|^2
\nonumber\\
&=
\frac{1}{MN}
\left|
\sum^{\,}_{m\in\mathcal{M}}
\sum^{N-1}_{i=0}
\langle m|i\rangle
\right|^2
=
\frac{M}{N}.
\label{eq: C function Grover asymptotic multiple}
\end{align}
We then obtain
$
\langle\Phi^{\,}_{0}|H^{\,}_{1}|\Phi^{\,}_{0}\rangle
=
1-
|\langle \Phi^{\,}_{1}|\Phi^{\,}_{0}\rangle|^{2}
=1-\frac{M}{N}.
$
Bringing this result to Eq.\ (\ref{eq: minimal runtime for projector Hamiltonian}) yields
$
T^{\,}_{\mathrm{inf}}=\mathcal{O}(\sqrt{N/M}).
$

\section{Detail calculation for adiabatic algorithm of finding k-clique in random graphs}
\label{sec: App 3}

We want to calculate $\langle\Phi^{\,}_{0}|H^{\,}_{1}|\Phi^{\,}_{0}\rangle\=:\overline{h^{\,}_{\mathrm{C}}}$
and $\langle\Phi^{\,}_{0}|H^{2}_{1}|\Phi^{\,}_{0}\rangle\=:\overline{h^{2}_{\mathrm{C}}},$
where
\begin{equation}
\begin{split}
&\overline{h^{\,}_{\mathrm{C}}}
=
\frac{1}{C(n,k)}
\sum^{\,}_{z\in\{0,1\}^n:|z|=k}
h^{\,}_{\mathrm{C}}(z),
\\
&
\overline{h^{2}_{\mathrm{C}}}
=
\frac{1}{C(n,k)}
\sum^{\,}_{z\in\{0,1\}^n:|z|=k}
h^{2}_{\mathrm{C}}(z),
\label{eq: define average of moments in appendix}
\end{split}
\end{equation}
with the cost function
$
h^{\,}_{\mathrm{C}}(z)=\sum^{\,}_{i>j}(1-G^{\,}_{ij})z^{\,}_{i}z^{\,}_{j}
$
as defined in Eq.\ (\ref{eq: final Hamiltonian adiabatic clique in RG eg equation}).
To conduct the calculation, one encounters terms like $\sum^{\,}_{i>j}G^{\,}_{ij},$ 
which cannot be done generically without knowing the explicit form of $G^{\,}_{ij}.$
We attempt a ``mean-field'' approach by considering the 
``average property'' of $G^{\,}_{ij}$ since each pair of vertices has probability $p$ to be connected 
and probability ($1-p$) otherwise. 
Note that the value of $p$ is $1/2$ in the main text.

\subsection{Mean-field approach}
\label{sec: App 3A}

We shall approximate $G^{\,}_{ij}$ by its expected value, $\mathbb{E}[G^{\,}_{ij}].$
Specifically, since each $G^{\,}_{ij}=1$ with probability $p$ and  $G^{\,}_{ij}=0$ with probability $1-p,$
we obtain
\begin{align}
\mathbb{E}[G^{\,}_{ij}]=p.
\label{eq: average of G}
\end{align}
We now compute $\overline{h^{\,}_{\mathrm{C}}}$ (\ref{eq: define average of moments in appendix}):
\begin{align}
\overline{h^{\,}_{\mathrm{C}}}
&=
\frac{1}{C(n,k)}
\sum^{\,}_{z\in\{0,1\}^n:|z|=k}
\sum^{\,}_{i>j}(1-G^{\,}_{ij})z^{\,}_{i}z^{\,}_{j}
\nonumber\\
&\approx
\mathbb{E}[\overline{h^{\,}_{\mathrm{C}}}]
\nonumber\\
&=
\frac{1}{C(n,k)}
\sum^{\,}_{z\in\{0,1\}^n:|z|=k}
\sum^{\,}_{i>j}(1-\mathbb{E}[G^{\,}_{ij}])z^{\,}_{i}z^{\,}_{j}.
\label{eq: uncertainty of driving term k clique in RG first term before}
\end{align}
There are $C(n,k)$ states $|z\rangle$ satisfy $|z|=k$ that we have to consider.
For each such state, there are $k$'s $z^{\,}_{i}$ for which $z^{\,}_{i}=1.$
Therefore, the first term in the parenthesis of Eq.\ (\ref{eq: uncertainty of driving term k clique in RG first term before}) is simply
\begin{align}
&\sum^{\,}_{z\in\{0,1\}^n:|z|=k}
\sum^{\,}_{i> j}
z^{\,}_{i}z^{\,}_{j}
=
\sum^{\,}_{z\in\{0,1\}^n:|z|=k}
C(k,2)
\nonumber\\
&=
C(n,k)\,C(k,2)
=
C(n,k)\,L^{\,}_{k}.
\label{eq: uncertainty of driving term k clique in RG first term before first term}
\end{align}
Here, for notational convenience, we define $L^{\,}_{k}\equiv C(k,2).$
The second term in the parenthesis of Eq.\ (\ref{eq: uncertainty of driving term k clique in RG first term before}) 
can be calculated with the help of Eq.\ (\ref{eq: average of G})
\begin{align}
\sum^{\,}_{z\in\{0,1\}^n:|z|=k}
\sum^{\,}_{i> j}
\mathbb{E}[G^{\,}_{ij}]
z^{\,}_{i}z^{\,}_{j}
=
p
C(n,k)L^{\,}_{k}.
\label{eq: uncertainty of driving term k clique in RG first term before second term}
\end{align}
Bringing Eqs.\ (\ref{eq: uncertainty of driving term k clique in RG first term before first term}) 
and (\ref{eq: uncertainty of driving term k clique in RG first term before second term}) back to 
Eq.\ (\ref{eq: uncertainty of driving term k clique in RG first term before}) yields
\begin{align}
\mathbb{E}[\overline{h^{\,}_{\mathrm{C}}}]
&=
\left(1-p\right)L^{\,}_{k}.
\label{eq: average of cost function from the first approach}
\end{align}

Next, we compute $\overline{h^{2}_{\mathrm{C}}}$ (\ref{eq: define average of moments in appendix}):
\begin{widetext}
\begin{align}
\overline{h^{2}_{\mathrm{C}}}
&=
\frac{1}{C(n,k)}
\sum^{\,}_{z\in\{0,1\}^n:|z|=k}
\sum^{\,}_{i'> j'}
\sum^{\,}_{i> j}
\left(
1-2G^{\,}_{ij}
+G^{\,}_{i'j'}G^{\,}_{ij}
\right)
z^{\,}_{i'}z^{\,}_{j'}
z^{\,}_{i}z^{\,}_{j}
\nonumber\\
&\approx
\mathbb{E}[\overline{h^{2}_{\mathrm{C}}}]
\nonumber\\
&=
\frac{1}{C(n,k)}
\sum^{\,}_{z\in\{0,1\}^n:|z|=k}
\sum^{\,}_{i'> j'}
\sum^{\,}_{i> j}
\Big(
1-2\mathbb{E}[G^{\,}_{ij}]
+\mathbb{E}\left[G^{\,}_{i'j'}G^{\,}_{ij}\right]
\Big)
z^{\,}_{i'}z^{\,}_{j'}
z^{\,}_{i}z^{\,}_{j}.
\label{eq: uncertainty of driving term p clique in RG second term before}
\end{align}
There are three terms in the parenthesis of Eq.\ (\ref{eq: uncertainty of driving term p clique in RG second term before}).
The first term is deterministic,
\begin{align}
\sum^{\,}_{z\in\{0,1\}^n:|z|=k}
\sum^{\,}_{i'> j'}
\sum^{\,}_{i> j}
z^{\,}_{i'}z^{\,}_{j'}
z^{\,}_{i}z^{\,}_{j}
&=
\sum^{\,}_{z\in\{0,1\}^n:|z|=k}
\left(
\sum^{\,}_{i'> j'}
z^{\,}_{i'}z^{\,}_{j'}
\right)
\left(
\sum^{\,}_{i> j}
z^{\,}_{i}z^{\,}_{j}
\right)
=
\sum^{\,}_{z\in\{0,1\}^n:|z|=k}
L^{\,}_{k}
L^{\,}_{k}
=
C(n,k)\,
L^{2}_{k}.
\end{align}
The second term in the parenthesis of Eq.\ (\ref{eq: uncertainty of driving term p clique in RG second term before}) reads
\begin{align}
\sum^{\,}_{z\in\{0,1\}^n:|z|=k}
\sum^{\,}_{i'> j'}
\sum^{\,}_{i> j}
\left(
-2
\mathbb{E}[G^{\,}_{ij}]
\right)
z^{\,}_{i'}z^{\,}_{j'}
z^{\,}_{i}z^{\,}_{j}
&=
-2
\sum^{\,}_{z\in\{0,1\}^n:|z|=k}
\left(
\sum^{\,}_{i'> j'}
z^{\,}_{i'}z^{\,}_{j'}
\right)
\left(
\sum^{\,}_{i> j}
\mathbb{E}[G^{\,}_{ij}]
z^{\,}_{i}z^{\,}_{j}
\right)
\nonumber\\
&\stackrel{(\ref{eq: average of G})}{=}
-2
\sum^{\,}_{z\in\{0,1\}^n:|z|=k}
\left(
\sum^{\,}_{i'> j'}
z^{\,}_{i'}z^{\,}_{j'}
\right)
\left(
\sum^{\,}_{i> j}
p
z^{\,}_{i}z^{\,}_{j}
\right)
=
-2p
C(n,k)
L^{2}_{k}.
\end{align}
Before proceeding to calculate the third term in the parenthesis of Eq.\ (\ref{eq: uncertainty of driving term p clique in RG second term before}),
we notice the following decomposition property
\begin{align}
\sum^{\,}_{i'> j'}
\sum^{\,}_{i> j}
\mathbb{E}\left[G^{\,}_{i'j'} G^{\,}_{ij}\right]
&=
\sum^{\,}_{i> j}
\mathbb{E}\left[G^{\,}_{ij} G^{\,}_{ij}\right]
+
\sum^{\,}_{i'> j',i> j:i\neq i',j\neq j'}
\mathbb{E}\left[G^{\,}_{i'j'} G^{\,}_{ij}\right]
\nonumber\\
&=
\sum^{\,}_{i> j}
\mathbb{E}\left[G^{2}_{ij}\right]
+
\sum^{\,}_{i'> j',i> j:i\neq i',j\neq j'}
\mathbb{E}\left[G^{\,}_{i'j'}\right]\mathbb{E}[G^{\,}_{ij}]
\nonumber\\
&=
\sum^{\,}_{i> j}
p
+
\sum^{\,}_{i'> j',i> j:i\neq i',j\neq j'}
p^2
=
L^{\,}_{k}p
+
\left(
L^{2}_{k}-L^{\,}_{k}
\right)
p^2.
\label{eq: average of G square}
\end{align}
Finally, the third term in the parenthesis of Eq.\ (\ref{eq: uncertainty of driving term p clique in RG second term before}) is
\begin{align}
\sum^{\,}_{z\in\{0,1\}^n:|z|=k}
\left(
\sum^{\,}_{i'> j'}
\sum^{\,}_{i> j}
\mathbb{E}\left[G^{\,}_{i'j'} G^{\,}_{ij}\right]
z^{\,}_{i'}z^{\,}_{j'}
z^{\,}_{i}z^{\,}_{j}
\right)
&\stackrel{(\ref{eq: average of G square})}{=}
p
C(n,k)
L^{2}_{k}
\left(
p
+
\frac{(1-p)}{L^{\,}_{k}}
\right).
\end{align}
\end{widetext}
Upon using the above results,
Eq.\ (\ref{eq: uncertainty of driving term p clique in RG second term before})
reads
\begin{align}
\mathbb{E}[\overline{h^{2}_{\mathrm{C}}}]
&=
\left(
1-2p+p\left(p+\frac{1-p}{L^{\,}_{k}}\right)
\right)
L^{2}_{k}
\nonumber\\
&=
(1-p)^2L^{2}_{k}+p(1-p)L^{\,}_{k}.
\label{eq: average of cost function square from the first approach}
\end{align}
Hence, the quantum uncertainty reads
$
\delta V^{\,}_{n}
=
\sqrt{p(1-p)L^{\,}_{k}}.
$
It then follows from Eq.\ (\ref{eq: lower bound on adiabatic runtime asymptotic}) a lower bound on runtime
\begin{align}
T^{\,}_{\mathrm{inf}}=
\mathcal{O}
\left(
\sqrt{1/L^{\,}_{k}}
\right),
\label{eq: runtime of finding k clique using approach one usual normalization}
\end{align}
which is independent of $n.$
Note that if the cost function $h^{\,}_{\mathrm{C}}(z)$ is rescaled as $h^{\,}_{\mathrm{C}}(z)\to h^{\,}_{\mathrm{C}}(z)/L^{\,}_{k},$
one should find 
$
T^{\,}_{\mathrm{inf}}=
\mathcal{O}
\left(
\sqrt{L^{\,}_{k}}
\right).
$

\subsection{Combinatorial approach}
\label{sec: App 3B}

In this section, we use an explicit combinatorial method to reproduce the same result as we found in 
Eq.\ (\ref{eq: runtime of finding k clique using approach one usual normalization}).
The following observation is crucial:
If a binary string $z$ of Hamming weight $k$ represents a graph that is a $k$-clique after adding $\alpha$ connected edges,
then $h^{\,}_{\mathrm{C}}(z)=\alpha,$ where $\alpha\in\{0,1,\cdots,C(k,2)\}.$

To simplify the notation, let us define $L^{\,}_{k}\equiv C(k,2)$ for convenience.
Now the question is what is the multiplicity $m^{\,}_{\alpha}$ for each possible value of cost function $h^{\,}_{\mathrm{C}}(z)=\alpha$.
Since each of the $L^{\,}_{k}$ edges has probability $p$ to be present when random graphs are generated uniformly, we expect
\begin{align}
m^{\,}_{\alpha}=p^{L^{\,}_{k}-\alpha}(1-p)^\alpha C(L^{\,}_{k},\alpha)\,C(n,k),
\end{align}
for $\alpha\in\{0,1,\cdots,L^{\,}_{k}\}$.
One verifies that the sum of all multiplicity equals $C(n,k),$ i.e., the number of binary strings of Hamming weight $k$
\begin{align}
\sum^{L^{\,}_{k}}_{\alpha=0}
m^{\,}_{\alpha}
&=
C(n,k)
\underbrace{
\sum^{L^{\,}_{k}}_{\alpha=0}
p^{L^{\,}_{k}-\alpha}(1-p)^\alpha 
C(L^{\,}_{k},\alpha)
}^{\,}_{=(p+1-p)^{L^{\,}_{k}}}
\nonumber\\
&=
C(n,k).
\end{align}

We proceed to compute $\mathbb{E}[\overline{h^{\,}_{\mathrm{C}}}]$:
\begin{align}
\mathbb{E}[\overline{h^{\,}_{\mathrm{C}}}]
&=\frac{1}{C(n,k)}\sum^{L^{\,}_{k}}_{\alpha=0}
m^{\,}_{\alpha}\alpha
=
\sum^{L^{\,}_{k}}_{\alpha=0}
p^{L^{\,}_{k}-\alpha}(1-p)^\alpha 
C(L^{\,}_{k},\alpha)\alpha
\nonumber\\
&=
(1-p)L^{\,}_{k},
\end{align}
which is the same as Eq.\ (\ref{eq: average of cost function from the first approach}).
We then compute $\mathbb{E}[\overline{h^{2}_{\mathrm{C}}}]$:
\begin{align}
\mathbb{E}[\overline{h^{2}_{\mathrm{C}}}]
&=\frac{1}{C(n,k)}\sum^{L^{\,}_{k}}_{\alpha=0}
m^{\,}_{\alpha}\alpha^2
=
\sum^{L^{\,}_{k}}_{\alpha=0}
p^{L^{\,}_{k}-\alpha}(1-p)^\alpha 
C(L^{\,}_{k},\alpha)\alpha^2
\nonumber\\
&=
(1-p)^2L^{2}_{k}+p(1-p)L^{\,}_{k},
\end{align}
which is identical to Eq.\ (\ref{eq: average of cost function square from the first approach}).


\bibliography{references-AQC}

\end{document}